\documentclass[prl,twocolumn,floatfix,showpacs,10pt,a4paper]{revtex4}
\usepackage{amssymb}
\usepackage{amsmath}
\usepackage[dvipdfm]{graphicx}
\usepackage{subfigure}
\usepackage{times}\usepackage[latin1]{inputenc}
\usepackage[all]{xy}
\usepackage{graphicx}
\newcommand{\bb}{\bibitem}
\newcommand{\be}{\begin{equation}}
\newcommand{\ee}{\end{equation}}
\newcommand{\ben}{\begin{eqnarray}}
\newcommand{\een}{\end{eqnarray}}
\newcommand{\bes}{\begin{subequations}}
\newcommand{\ees}{\end{subequations}}

\begin{document}
\title{Bright and Dark Solitons in a Periodically Attractive and Expulsive Potential\\
with Nonlinearities Modulated in Space and Time}
\author{A.T. Avelar$\,^a$, D. Bazeia$\,^b$, and W.B. Cardoso$\,^a$}
\affiliation{$^a$Instituto de F\'\i sica, Universidade Federal de Goi\'as, 74.001-970, Goi\^ania
Goi\'as, Brazil\\$^b$Departamento de F\'\i sica, Universidade Federal da Para\'\i ba, 58.051-970, Jo\~ao
Pessoa Para\'\i ba, Brazil}
%%%%%%%%%%%%%%%%%%%%%%%%%%%%%%%%%%%%%%%%%%%%%%%%%%%%%%%
\begin{abstract}
This work deals with soliton solutions of the nonlinear Schroedinger equation with a diversity of nonlinearities. We solve the equation in a potential which oscillates in time between attractive and expulsive behavior, in the presence of nonlinearities which are modulated in space and time. Despite the presence of the periodically expulsive behavior of the potential, the results show that the nonlinear equation can support a diversity of localized excitations of the bright and dark type which are remarkably robust.
\end{abstract}

\maketitle

%%%%%%%%%%%%%%%%%%%
%%%%%%%%%%%%%%%%%%%
Solitons are known to appear in the nonlinear Schroedinger equation (NLSE) in a diversity of situations. Usually, the nonlinear interactions are of cubic nature, but there are systems \cite {B1,B2,B3} which engender cubic, quintic, cubic plus quintic, and other forms of nonlinearities of current interest to the study of Bose-Einstein condensates (BECs). The presence of BEC of trapped atoms was experimentally realized in \cite{E1,E2,E3}. Not too much later, dark solitons in BEC were formed with repulsive $^{78}$Rb atoms in \cite{DS1,DS2}, and bright solitons were generated for attractive $^7$Li atoms in \cite{BS1,BS2}. Along the years, many interesting advancements and motivations have appeared, as we can see, for instance, in Refs.~\cite{B4,B5}.

There are several distinct lines of investigation of the NLSE. An interesting possibility was initiated in \cite{SH}, with the introduction of a procedure to deal with the NLSE with varying coefficients. It is based on a similarity transformation, which transforms nonautonomous NLSE into stationary equation which is easier to solve \cite{SH,SHB1}. In particular, in the recent work \cite{BB} one deals with the cubic NLSE where both the trapping potential and the cubic nonlinearity may now vary, being functions of both space and time. This procedure was recently extended to the case of cubic and quintic nonlinearities in \cite{abc}.

Another line of investigation of the NLSE was initiated in \cite{sci}, in which one considers the possibility of changing the potential from the usual attractive harmonic form to the well distinct expulsive behavior, tuned through the Feshbach resonance mechanism. This possibility has initiated several investigations with expulsive potentials, with a variety of motivations of current interest \cite{cc,W,lzl,MM}.

Since it is believed that atomic matter nonlinear excitations are of importance for the development of applications of BECs, we think it is of interest to develop new procedures which allow us to investigate the presence of bright and dark solitons in realistic models. With this motivation, in this Letter we shall join together the two possibilities above, that is, we shall consider the NLSE with space- and time-dependent nonlinearities in the presence of space-dependent potential which oscillates periodically in time from attractive to expulsive behavior. The main results are that despite the expulsive behavior of the potential, the system is still able to support soliton excitations of both the bright and dark type, with a diversity of specific features, driven by the parameters which specify the system. The results are robust since the solutions run in time with periodic or quasiperiodic features without loosing the localized behavior in space.

We shall implement the procedure focusing attention on the NLSE of the general form
\begin{equation}
i\frac{\partial\Psi}{\partial t}\!=\!-\frac{\partial^{2}\Psi}{\partial x^{2}}+v\!(x,t)\Psi\!+P(|\Psi|^2)\Psi, \label{NLSE}
\end{equation}
where $v(x,t)$ is the potential and $P(|\Psi|^2)$ is the polynomial function
\be
P(|\Psi|^2)=\sum_{n=1}^N g_{2n+1}(x,t)|\Psi|^{2n}.
\ee
The coefficients $g_{2n+1}(x,t)$ describe the nonlinearities present in the system. Here we are using standard notation, in which $x$, $t$ and all other quantities are dimensionless.

We assume validity of the one dimensional approach, which is an approximation of good practical use if one works under the restrictions $a_s\ll a_{\bot}\ll\lambda\ll a_{||}$, for $a_s$ being the scattering length, $\lambda$ the spatial scale of the wave packet, and $a_{\bot}$ and $a_{||}$ being the characteristics transverse and longitudinal trap lengths, respectively. We also remark that issues concerning parametric amplification of elementary excitations due to periodic modulation of the trapping potential and nonlinearities were studied before in \cite{SLV}. Moreover, the potential and nonlinearities to be used in this work are typical of BECs, so that the results obtained below may stimulate new experiments in the field.

%%%%%%%%%%%%%%%%%%%%
%%%%%%%%%%%%%%%%%%%%
The main idea of the procedure is to transform the nonautonomous NLSE \eqref{NLSE} into the stationary equation
\be
\mu\Phi(\zeta )=-\frac{d^{2}\Phi(\zeta)}{d\zeta^{2}}+Q(|\Phi|^2)\Phi(\zeta),\label{Q}
\ee
where $\mu$ is the eigenvalue of the nonlinear equation above, and $Q(|\Phi|^2)$ has the form
\be
Q(|\Phi|^2)=\sum_{n=1}^N G_{2n+1}|\Phi|^{2n},
\ee
where the coefficients $G_{2n+1}$ are now real and constant parameters.
To do this, we write the solution of \eqref{NLSE} as \cite{SH,BB}
\begin{equation}
\Psi (x,t)=\rho(x,t)e^{i\phi(x,t)}\Phi(\zeta(x,t)),\label{TS}
\end{equation} 
The substitution of \eqref{TS} into \eqref{NLSE} leads to (\ref{Q}), but now we have to have 
\bes\label{nl1}\ben
&&\rho\frac{\partial\rho}{\partial t}+\frac{\partial(\rho^{2}(\partial\phi/\partial x))}{\partial x}=0,
\\
&&\frac{\partial\zeta}{\partial t}+2\frac{\partial\phi}{\partial x}\frac{\partial\zeta}{\partial x}=0,\;\;\;\;\;
\frac{\partial\left(\rho^{2}(\partial\zeta/\partial x)\right)}{\partial x}=0.
\een\ees

Next, we introduce a new function $\xi(x,t)$ such that $\zeta(x,t)=F(\xi(x,t))$. If we make the simple choice $\xi(x,t)=x/\chi(t)$, we can determine the width of the localized solution as $\chi(t)$. With this, we obtain the new equations
\bes\label{nl2}\ben
\rho (x,t)&=&\left({\chi}{(dF/d\xi)}\right)^{-1/2},
\\
\phi(x,t)&=&\frac1{4\chi}\,\frac{d\chi}{dt}\,x^{2},\label{q}
\\
v(x,t)&=&\frac{1}{\rho}\frac{\partial^{2}\rho}{\partial x^{2}}-\left(\frac{\partial\phi}{\partial x}\right)^{2}-\frac{\partial\phi}{\partial t}-
\frac{\mu}{\chi^4\rho^4},\label{vx}
\\
g_{2i+1}(x,t)&=&\frac{G_{2i+1}}{{\chi^{4}\rho^{4+2i}}},\;\;\;\;i=1,2,3,...\label{nl2b}
\een\ees
The choice of $F(\xi)$ is to be done consistently. 

To illustrate the procedure with examples of interest, let us now focus attention to the case of specific nonlinearities. There are many possibilities to choose the nonlinearities, and here we suppose that the cubic nonlinearity is given explicitly by  
\be\label{g3}
g_3(x,t)={\rm sgn}(G_3)\,\chi^{-1}\,[1+\lambda\exp(-\xi^2)]^{3} ,
\ee
where $\lambda$ is real parameter which controls the behavior of the cubic nonlinearity -- here we will use $\lambda=0.5$ where required. With this choice we have
\be
\rho(x,t)={\rm abs}(G_3)^{1/6}{\chi^{-1/2}[1+\lambda\exp(-\xi^2)]^{-1/2}},\label{ro}
\ee
and the function $F(\xi)$ has to be obtained from ${dF}/{d\xi}={\rm abs}(G_3)^{-1/3}[1+\lambda \exp(-\xi^2)]$.
The potential in Eq.~\eqref{vx} has to have the form
\ben\label{pot}
v(x,t)\!&=&\!\omega(t)x^{2}\!+\!f(x,t)\!-\!\mu\, \frac{[1\!+\!\lambda\exp(-\xi^2)]^2}{{\rm abs}(G_3)^{2/3}{\chi^{2}}},
\een
where $\omega$ is time-dependent function such that
\be
\omega(t)=-\frac{1}{4\chi}\frac{d^{2}\chi}{dt^{2}},\label{11a}
\ee
and $f(x,t)$ is an awkward expression which comes from the first term in \eqref{vx}. The other nonlinearities are given according to Eq.~\eqref{nl2b}.
We remark that the last two terms in the potential \eqref{pot} contribute most significantly at small distances, as we show explicitly below. 

There are several possibilities to choose the width of the localized excitation. In the present work we consider
\be\label{chi}
\chi(t)=\frac{1}{1+[1+\alpha\sin(t)+\beta\sin(\sqrt{2}t)]^2},
\ee
where $\alpha$ and $\beta$ are real parameters. This choice leads to $\omega(t)$ according to Eq.~\eqref{11a} and in Fig.~1 we plot $\omega(t)$
for $\alpha=0.1$ and $\beta=0$, and for $\alpha=\beta=0.1$, to illustrate its periodic and quasiperiodic features.

%%%%%%%%%%%%%%%%%%%%%%%%%%%%%%%%%%%%%%%%%%%%%%%%%%%%%%%%%%
\begin{figure}[t]
\includegraphics[width=8cm]{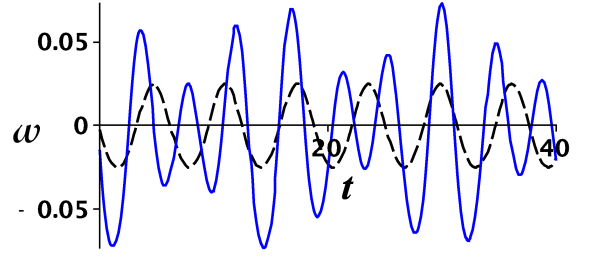}
\caption{(Color on line) Plots of $\omega(t)$ in black (dashed) and blue (solid) lines, for $\alpha=0.1$ and $\beta=0$, and for $\alpha=\beta=0.1$, respectively.}
\end{figure}
%%%%%%%%%%%%%%%%%%%%%%%%%%%%%%%%%%%%%%%%%%%%%%%%%%%%%%%%%%%

The above results are general and can be used to investigate systems of current interest. Let us first consider the NLSE in the case of cubic nonlinearity alone. This is the case of a BEC where the two-body scattering is strong enough to allow neglecting the other higher order effects. For instance, in a BEC system of atomic species with naturally large interactions such as $^{87}{\rm Rb}$ \cite{E1} and $^{23}{\rm Na}$ \cite{E3} have properties favorable to be described by the cubic nonlinearity above. This leads to the stationary equation 
\be \label{CTS1}
\mu\Phi(\zeta )=-\frac{d^{2}\Phi(\zeta)}{d\zeta^{2}} +G_3|\Phi(\zeta)|^{2}\Phi(\zeta).
\ee

We now explore the presence of bright and dark solitons. We first take $\mu=-1$ e $G_3=-2$ to get to the potential
\ben
v(x,t)\!\!=\!\omega(t)x^{2}\!\!+\!\!f(x,t)\!+\!2^{-2/3}\chi^{-2}[1\!+\!\lambda\exp(-\xi^2)]^2,\label{vcb}
\een
with $g_3(x,t)$ given by \eqref{g3}. In Fig.~2(a) we plot the potential \eqref{vcb} to show how it behaves in space and time. Moreover,
for $\mu=-1$ e $G_3=-2$ the equation \eqref{CTS1} has the solution
\be\label{solC1bs}
\Phi(\zeta)=\mathrm{sech}(\zeta),
\ee
which is of the bright type. The wave function which solves the corresponding NLSE gets to the form
\be\label{psicb}
\Psi(x,t)={2^{1/6}{e^{i\phi}}{\rm sech(\zeta)}}{\{\chi[1+\lambda\exp(-\xi^2)]\}^{-1/2}},
\ee
where $\phi=\phi(x,t)$ is real, obtained via the equation \eqref{q}.

To get to the case of dark solitons, we consider $\mu=2$ and $G_3=2$. In this case the potential becomes
\ben
v(x,t)\!\!=\!\omega(t)x^{2}\!\!+\!\!f(x,t)-2^{1/3}\chi^{-2}[1+\lambda\exp(-\xi^2)]^2.\label{vcd}
\een
In Fig.~2(b), we plot the potential \eqref{vcd} to illustrate how it behaves in space and time. If we use $\mu=2$ and $G_3=2$ in Eq.~\eqref{CTS1}, we get the solution
\be\label{phicd}
\Phi(\zeta)=\tanh(\zeta),
\ee
which is of the dark type. The wave function which solves the corresponding NLSE gets to the form
\be\label{psicd}
\Psi(x,t)={2^{1/6}{e^{i\phi}}\tanh(\zeta)}{\{\chi[1+\lambda\exp(-\xi^2)]\}^{-1/2}},
\ee
with $\phi=\phi(x,t)$ real, obtained via the equation \eqref{q}.

%%%%%%%%%%%%%%%%%%%%%%%%%%%%%%%%%%%%
\begin{figure}[t]
\includegraphics[width=4.0cm]{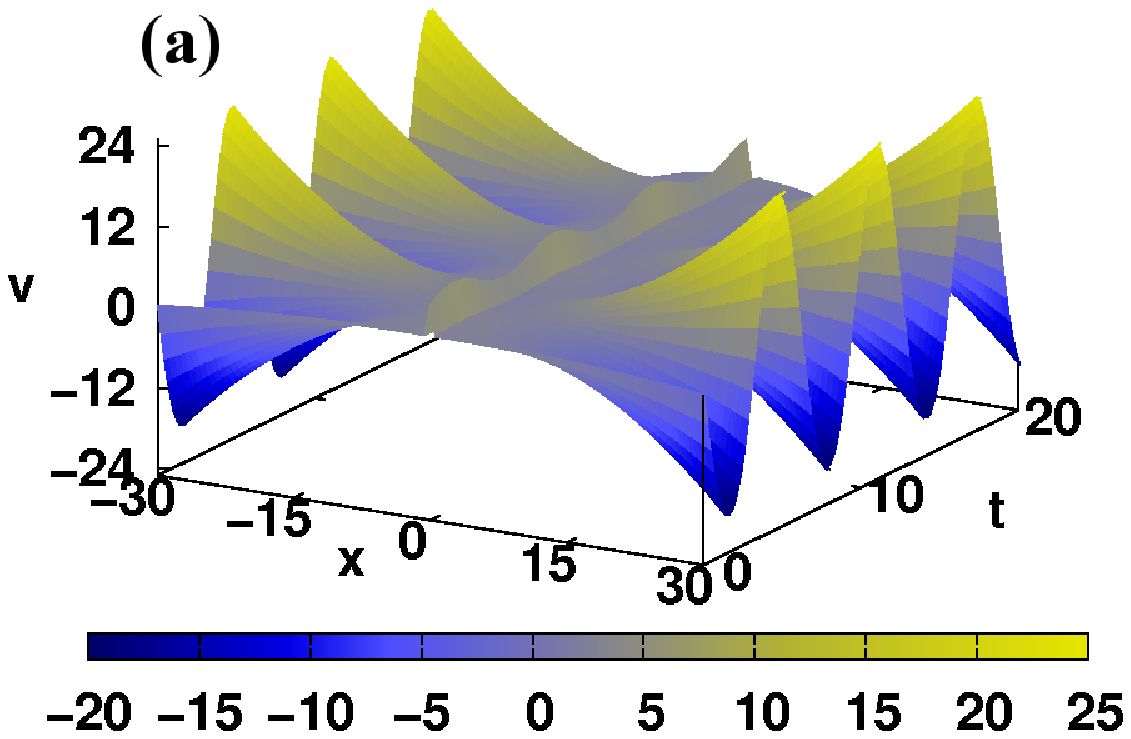}\hspace{0.5cm}
\includegraphics[width=4.0cm]{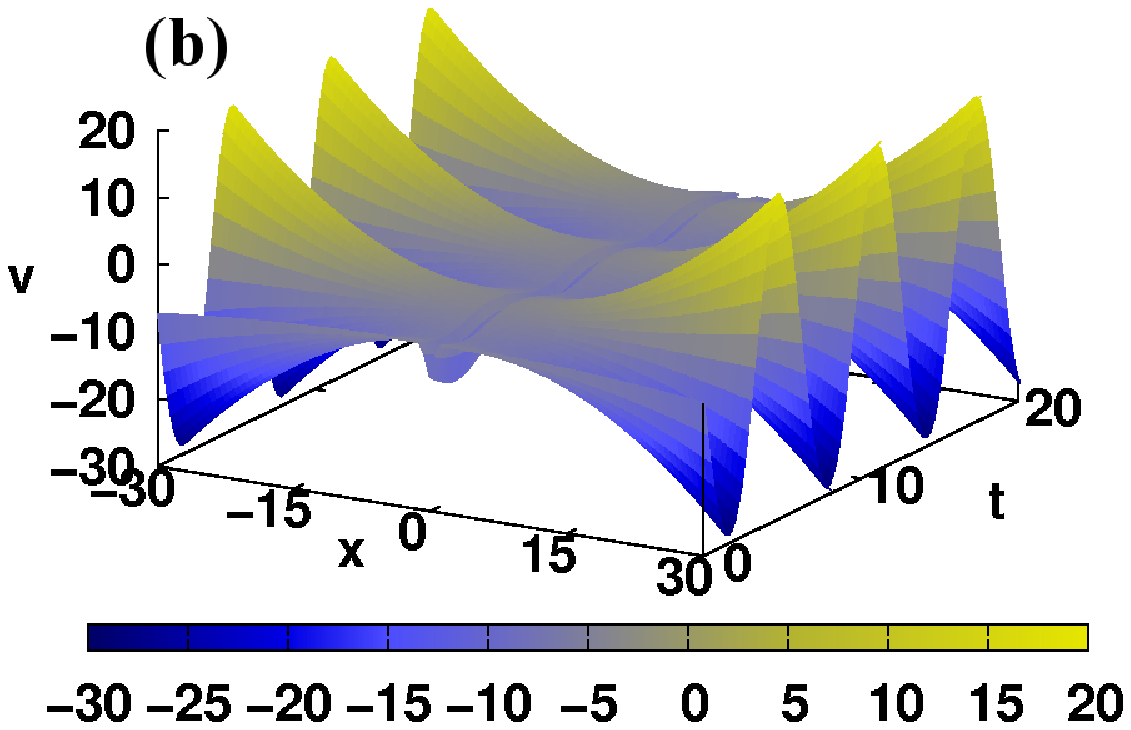}
\caption{(Color on line) Plots of $v(x,t)$ in the presence of cubic nonlinearity, with $\lambda=0.5$ and for $\omega(t)$ periodic, with $\alpha=0.1$ and $\beta=0$, as shown in Fig.~1. The plot (a) is of \eqref{vcb} for bright solitons, and the plot (b) is of \eqref{vcd} for dark solitons.}
\end{figure}
%%%%%%%%%%%%%%%%%%%%%%%%%%%%%%%%%%%%%%%%%%%%%

%%%%%%%%%%%%%%%%%%%%%%%%%%%%%%%%%%%%
\begin{figure}[t]
\includegraphics[width=4.0cm]{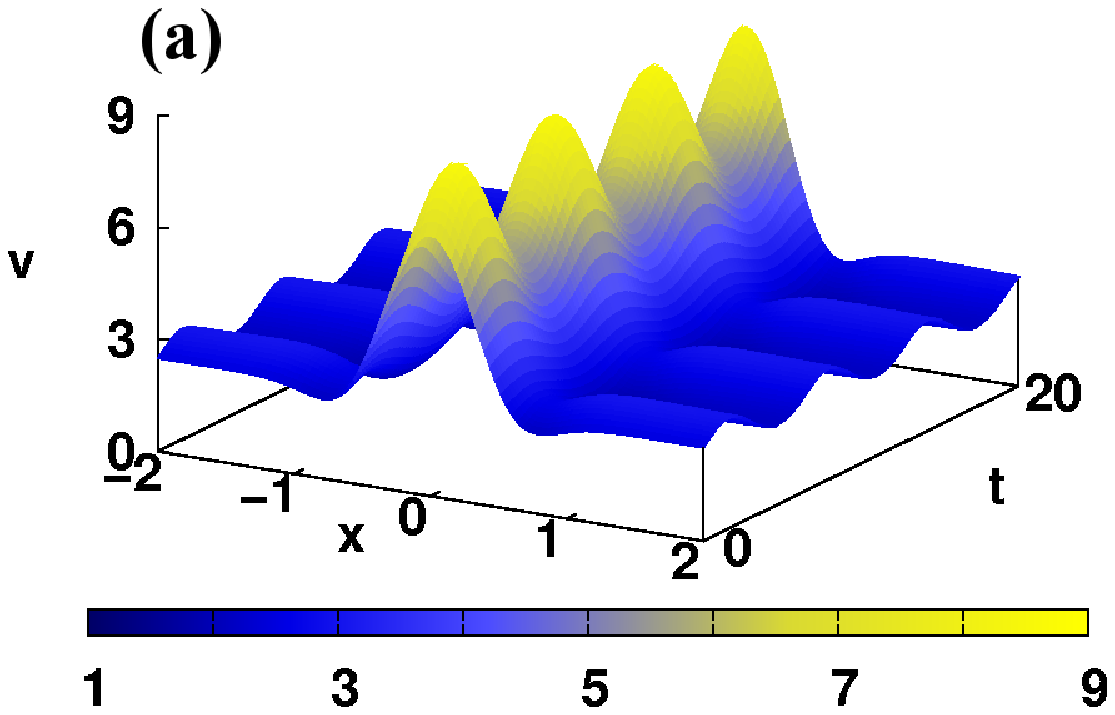}\hspace{0.5cm}
\includegraphics[width=4.0cm]{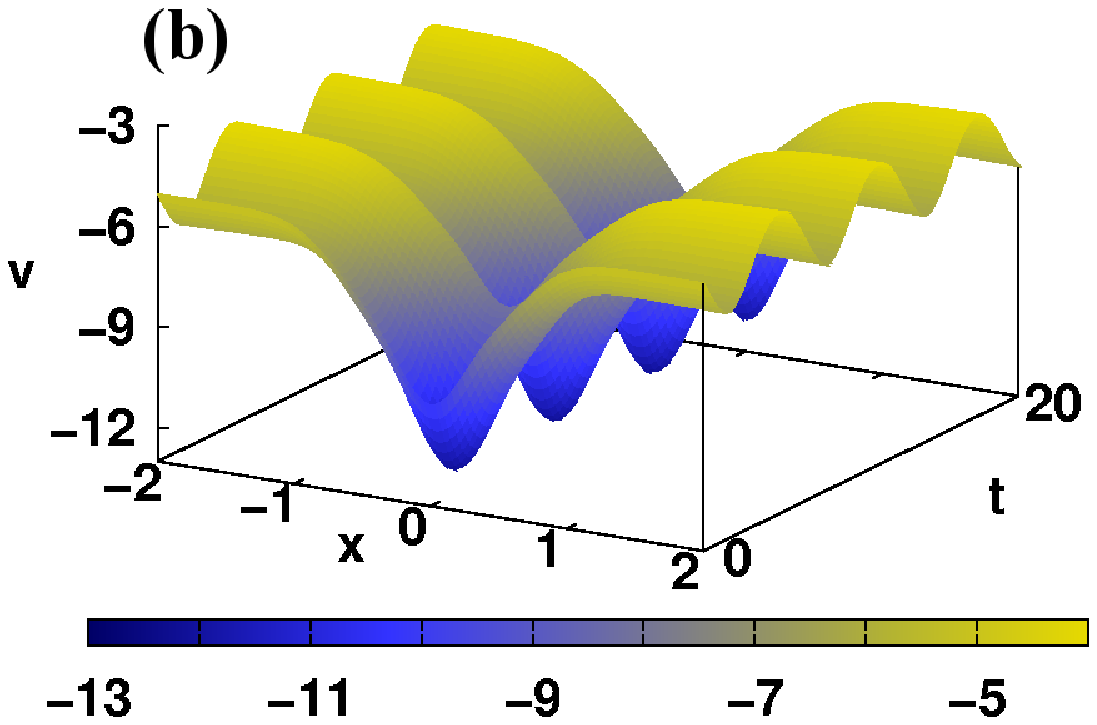}
\caption{(Color on line) Plots of $v(x,t)$ in the presence of cubic nonlinearity, as given in the former Fig.~2, but now in the range $-2<x<2$,
to expose the behavior of the potential at small distances.}
\end{figure}
%%%%%%%%%%%%%%%%%%%%%%%%%%%%%%%%%%%%%%%%%%%%%

We note from Fig.~2 that the potentials oscillate from attractive to expulsive behavior, and that they show some structure at small distances. To better see this, in Fig.~3 we plot the same potentials, but now we look at the behavior at small distances. We see that there are attractive and expulsive structures at small distances, but they do not contribute importantly to the soliton formation. This conclusion follows from the fact that if we change $\lambda\to-\lambda$ in \eqref{g3}, the structure at small distances changes significantly, but the soliton solutions maintain the same form, with changes mostly in the amplitude. The other potentials show similar behavior, and the issue will be further explored in another work \cite{abcNext}.

We now focus attention to the bright and dark solitons in the presence of cubic nonlinearity. In Figs.~4(a) and 4(b) we depict $|\Psi|^2$ according to \eqref{psicb}, for $\omega(t)$ periodic and quasiperiodic, respectively. In Figs.~5(a) and 5(b) we depict $|\Psi|^2$ according to \eqref{psicd}, for $\omega(t)$ periodic and quasiperiodic, respectively. We note that the solutions evolve in time periodically or quasiperiodically, depending on the way we choose $\chi(t)$ in \eqref{chi}, with no further changes, showing that they are stable localized excitations which we name periodic and quasiperiodic bright and dark solitons of the corresponding cubic NLSE.

%%%%%%%%%%%%%%%%%%%%%%%%%%%%%%%%%%%%%%%%%%%
\begin{figure}[t]
\includegraphics[width=4.1cm]{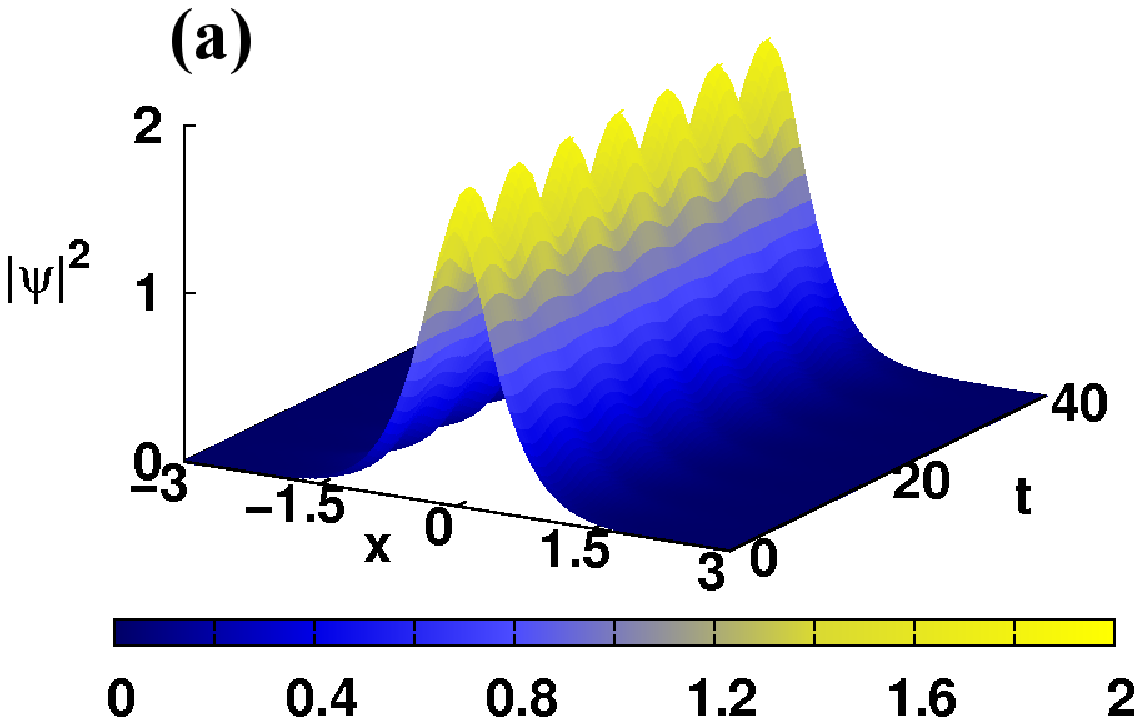}\hspace{0.2cm}
\includegraphics[width=4.1cm]{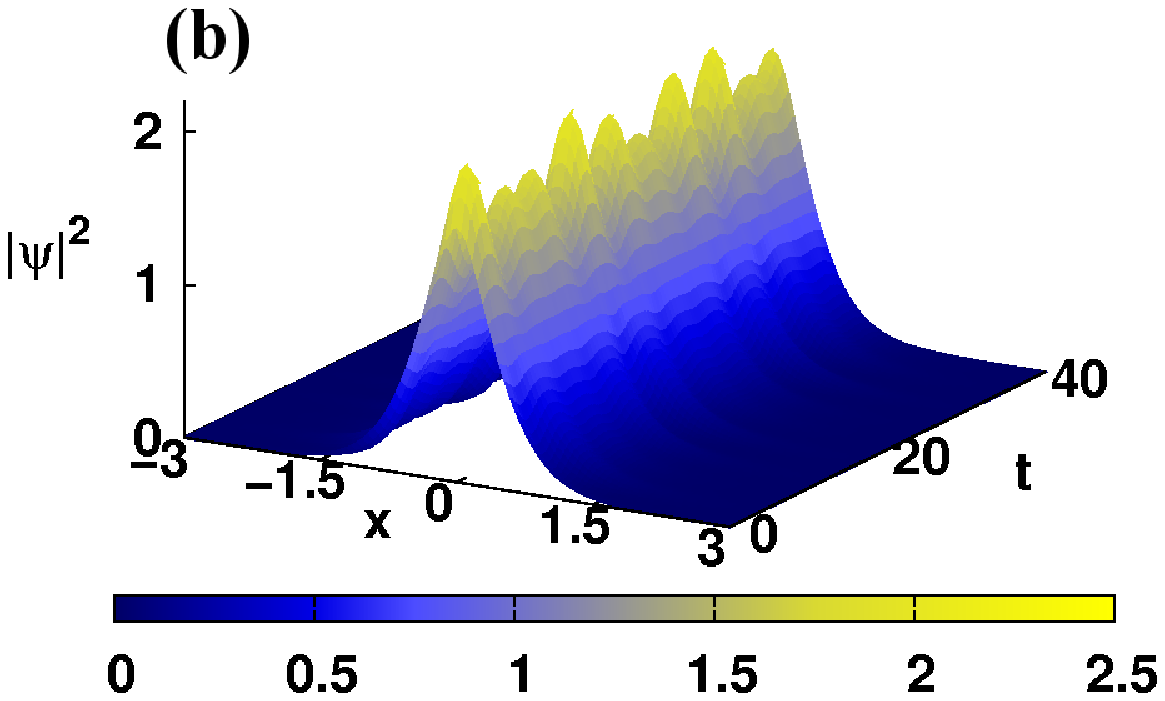}
\caption{(Color on line) Plots of $|\Psi(x,t)|^2$ with $\lambda=0.5$ for the periodic (a) or quasiperiodic (b) bright soliton, with $\alpha=0.1$ and $\beta=0$ or $\alpha=\beta=0.1$, respectively, in the case of cubic nonlinearity.}
\end{figure}
%%%%%%%%%%%%%%%%%%%%%%%%%%%%%%%%%%%%%%%%

%%%%%%%%%%%%%%%%%%%%%%%%%%%%%%%%%%%%%%%%%%%
\begin{figure}[t]
\includegraphics[width=4.1cm]{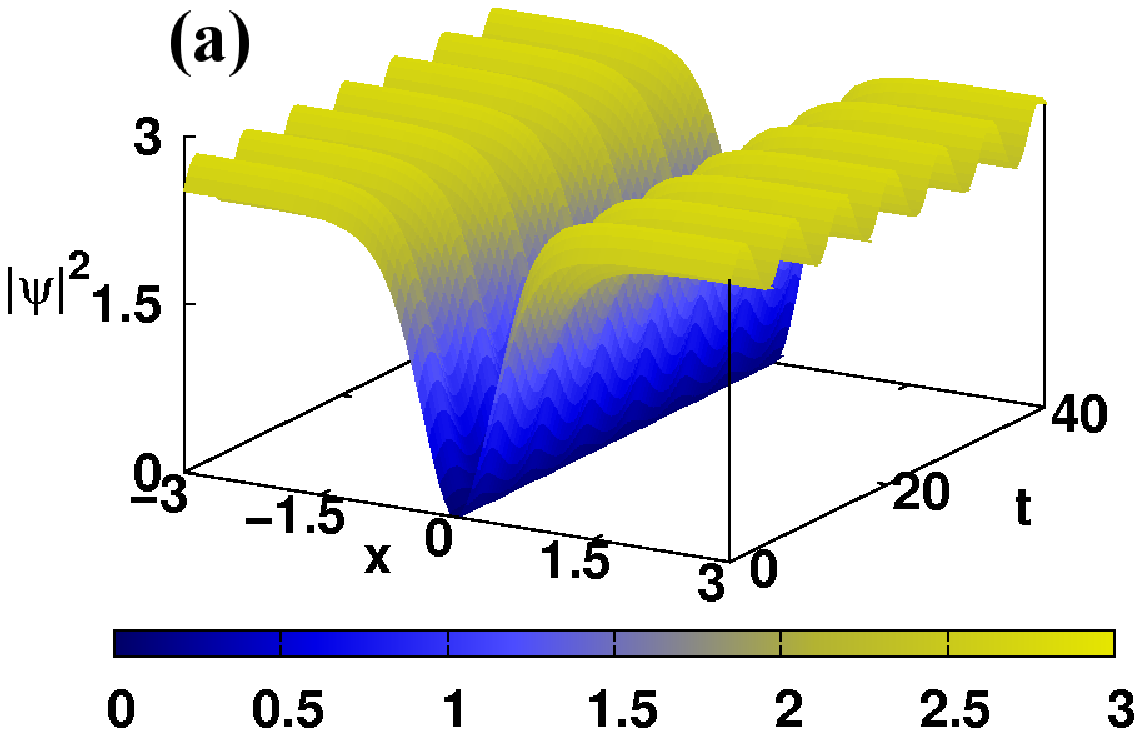}\hspace{0.2cm}
\includegraphics[width=4.1cm]{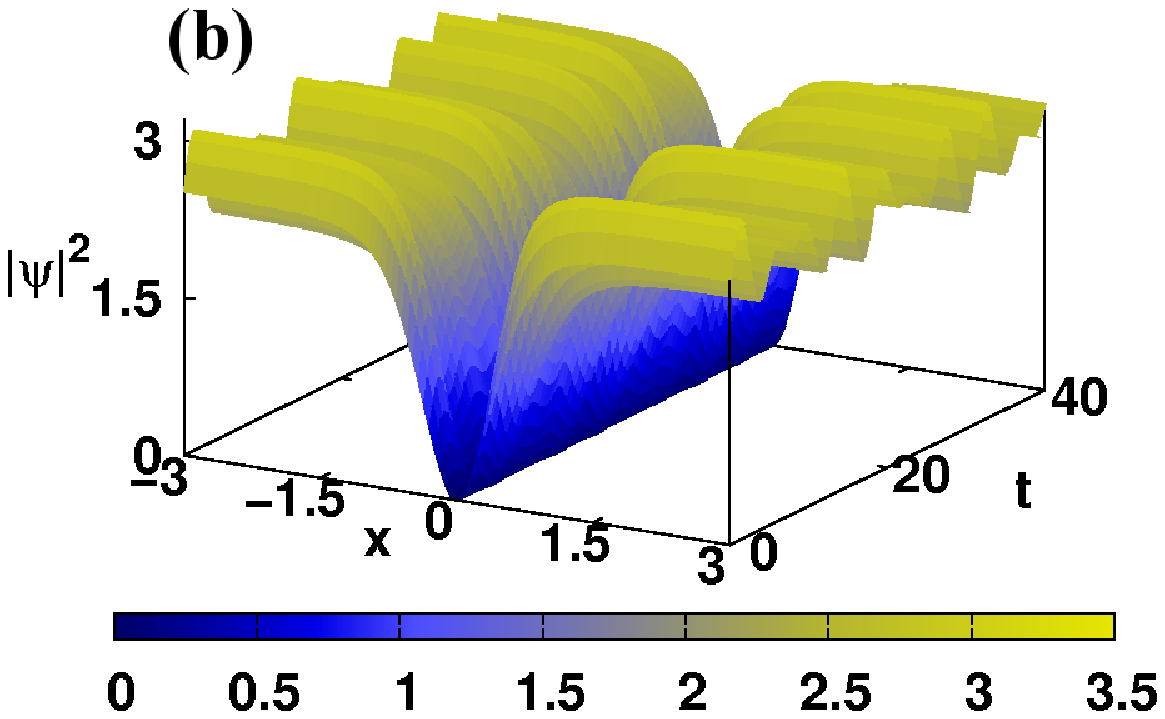}
\caption{(Color on line) Plots of $|\Psi(x,t)|^2$ with $\lambda=0.5$ for the periodic (a) or quasiperiodic (b) dark soliton, with $\alpha=0.1$ and $\beta=0$ or $\alpha=\beta=0.1$, respectively, in the case of cubic nonlinearity.}
\end{figure}
%%%%%%%%%%%%%%%%%%%%%%%%%%%%%%%%%%%%%%%%

%%%%%%%%%%%%%%%%%%%%%%%%%%%%%%%
%%%%%%%%%%%%%%%%%%%%%%%%%%%%%%%
We now consider the case of cubic and quintic nonlinearities. This case is also of interest, since the presence of quintic nonlinearity is in general due to
three-body interactions, and the cubic and quintic nonlinearities may be used to map the case where the two-body scattering is somehow weakened in a way such that the three-body effects cannot be neglected anymore \cite{G}. Moreover, it may be also used to help control deviation of the trapped condensate from
one-dimensionality \cite{kvm}. Here we use the same cubic nonlinearity \eqref{g3}, but now the quintic nonlinearity has the form
\be\label{g5}
g_5=G_5\,{\rm abs}(G_3)^{-4/3}[1+\lambda\exp(-\xi^2)]^4.
\ee
In this case, the function $\Phi(\zeta)$ obeys
\be\label{QC}
\mu\Phi(\zeta )=-\frac{d^{2}\Phi(\zeta)}{d\zeta^{2}}+G_3|\Phi(\zeta)|^{2}\Phi(\zeta)+G_5|\Phi(\zeta)|^{4}\Phi(\zeta).
\ee
The investigation is similar to the case considered in \cite{abc}. We first consider $\mu=0$, and take the cubic and quintic parameters as $G_3=2$ and $G_5=-3$. 
In this case we get the potential
\be
v(x,t)=\omega(t)x^{2}+f(x,t),\label{vcqb}
\ee
with $g_3(x,t)$ and $g_5(x,t)$ given by \eqref{g3} and \eqref{g5}, respectively. Moreover, for $\mu=0$, and for $G_3=2$ and $G_5=-3$, we can write Eq.~\eqref{Q} in the form 
\be
\frac{d^{2}\Phi(\zeta)}{d\zeta^{2}}=2\Phi^3(\zeta)-3\Phi^5(\zeta).\label{BS}
\ee
The solution is given by $\Phi(\zeta )={1}/{\sqrt{1+\zeta^{2}}}$. It has the bell shape form, and it is of the bright type.
In this case the wave function which solves the corresponding NLSE acquires the form
\be\label{psicqb}
\Psi(x,t)=2^{1/6}e^{i\phi}\{\chi[1+\lambda\exp(-\xi^2)](1+\zeta^2)\}^{-1/2},
\ee
where $\phi=\phi(x,t)$ is real, obtained via the Eq.~\eqref{q}.

To get to the darklike solution we choose, for instance, $\mu=3$, $G_3=6$, and $G_5=-3$. In this case the potential becomes
\be
v(x,t)\!=\!\omega(t)x^{2}\!+\!f(x,t)\!-\!(3/4)^{1/3}\chi^{-2}[1\!+\!\lambda\exp(-\xi^2)]^2,\label{vcqd}
\ee
with $g_3(x,t)$ and $g_5(x,t)$ given by \eqref{g3} and \eqref{g5}, respectively. Moreover, for $\mu=3$, and for $G_3=6$ and $G_5=-3$, we can use Eq.~\eqref{Q} to get to the  solution $\Phi(\zeta)={\zeta}/{\sqrt{1+\zeta^{2}}}$. It has the form of a kink, and so $|\Phi(\zeta)|^{2}$ is now of the dark type.
In this case, the solution given by Eq.~\eqref{TS} takes the form
\be\label{psicqd}
\Psi(x,t)=6^{1/6}e^{i\phi}\zeta\{\chi[1+\lambda\exp(-\xi^2)](1+\zeta^2)\}^{-1/2}.
\ee

%%%%%%%%%%%%%%%%%%%%%%%%%%%%%%%%%%%%
\begin{figure}[t]
\includegraphics[width=4.1cm]{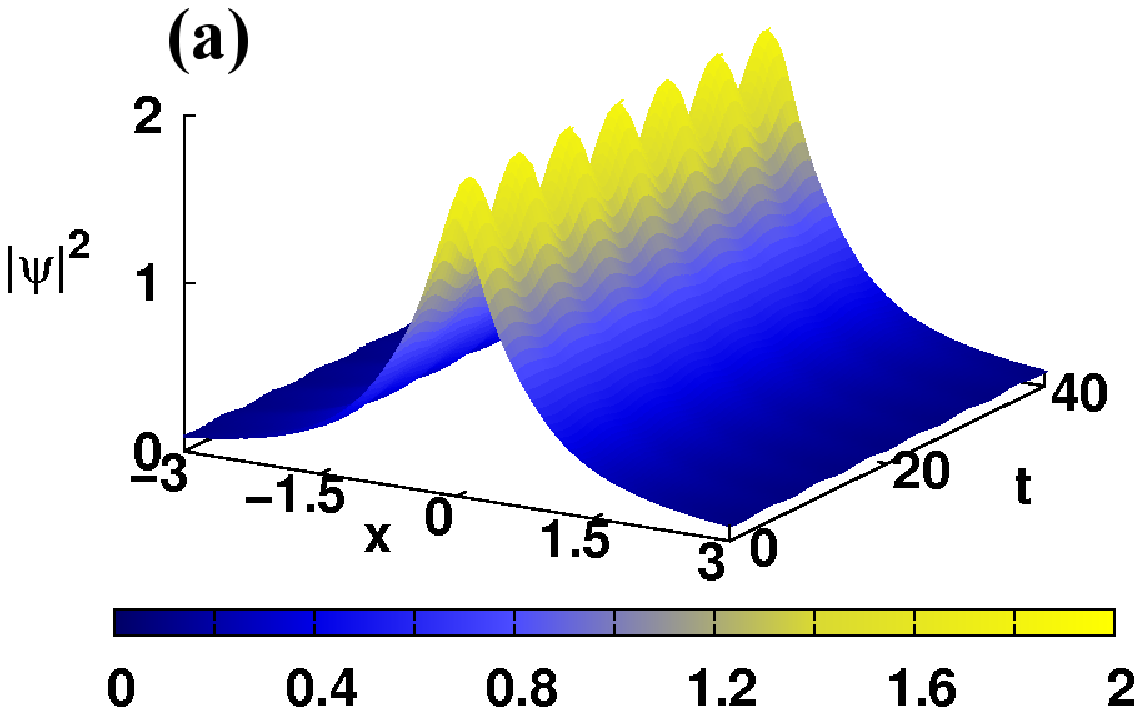}\hspace{0.2cm}
\includegraphics[width=4.1cm]{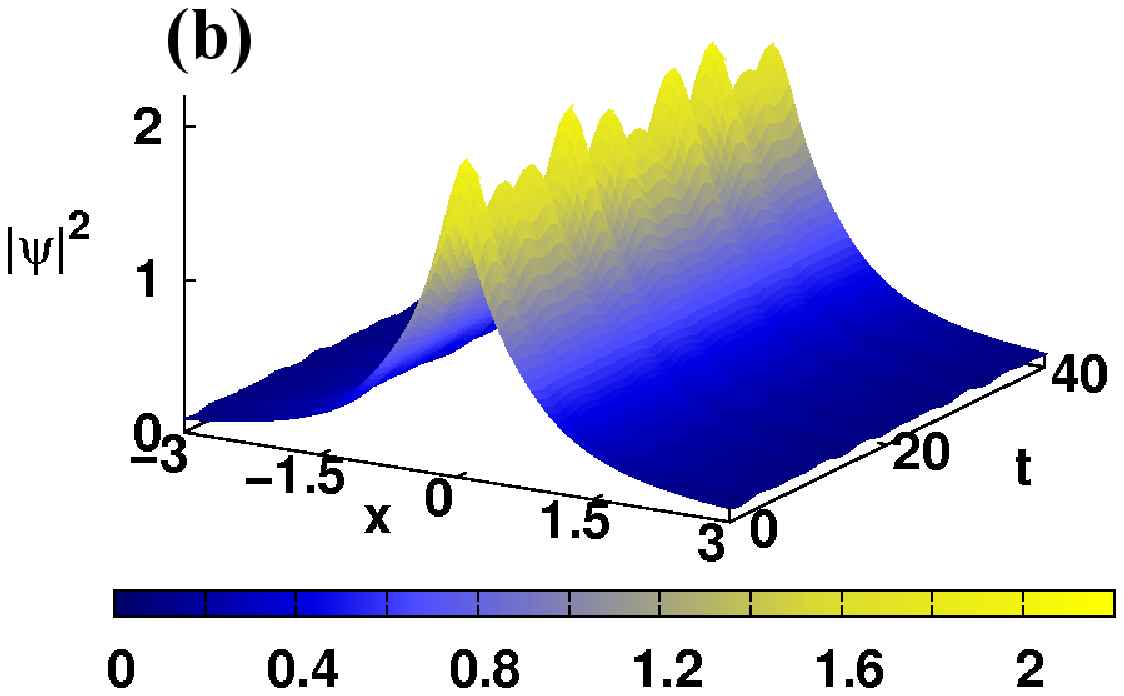}
\caption{(Color on line) Plots of $|\Psi(x,t)|^{2}$ with $\lambda=0.5$ for the periodic (a) or quasiperiodic (b) bright soliton, with $\alpha=0.1$ and $\beta=0$ or $\alpha=\beta=0.1$, respectively, in the case of cubic and quintic nonlinearities.}
\end{figure}
%%%%%%%%%%%%%%%%%%%%%%%%%%%%%%%
%%%%%%%%%%%%%%%%%%%%%%%%%%%%%%%%%%%%%%%%%%
\begin{figure}[t]
\includegraphics[width=4.1cm]{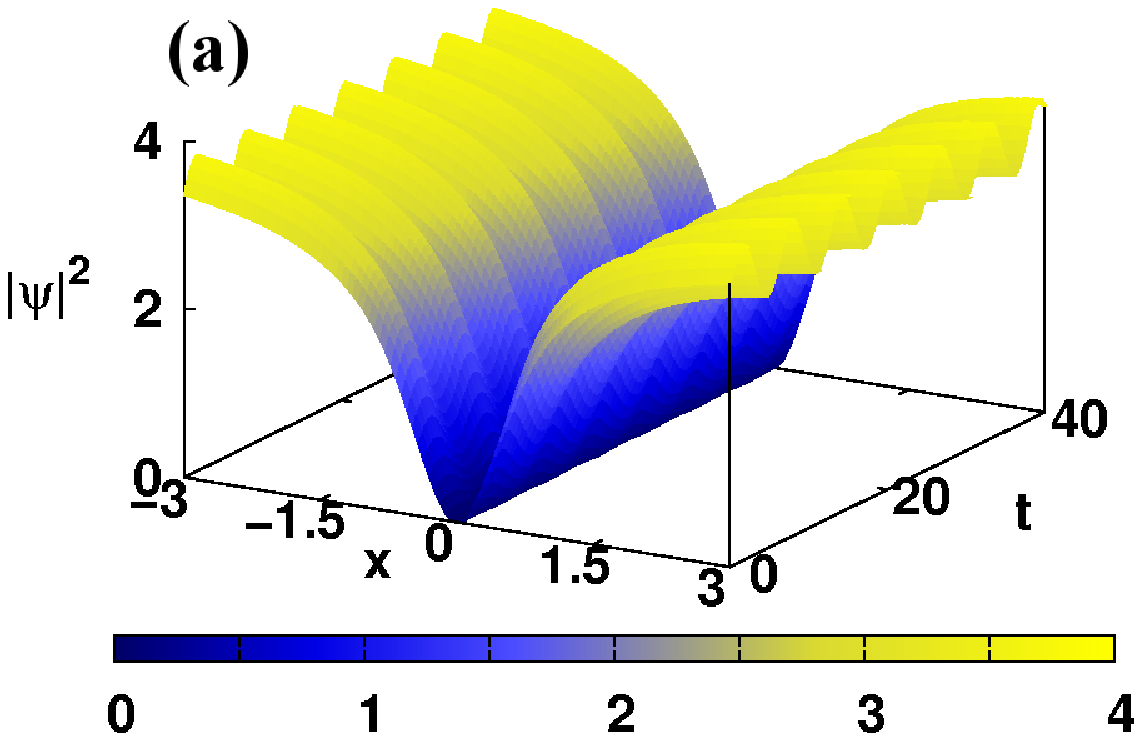}\hspace{0.2cm}
\includegraphics[width=4.1cm]{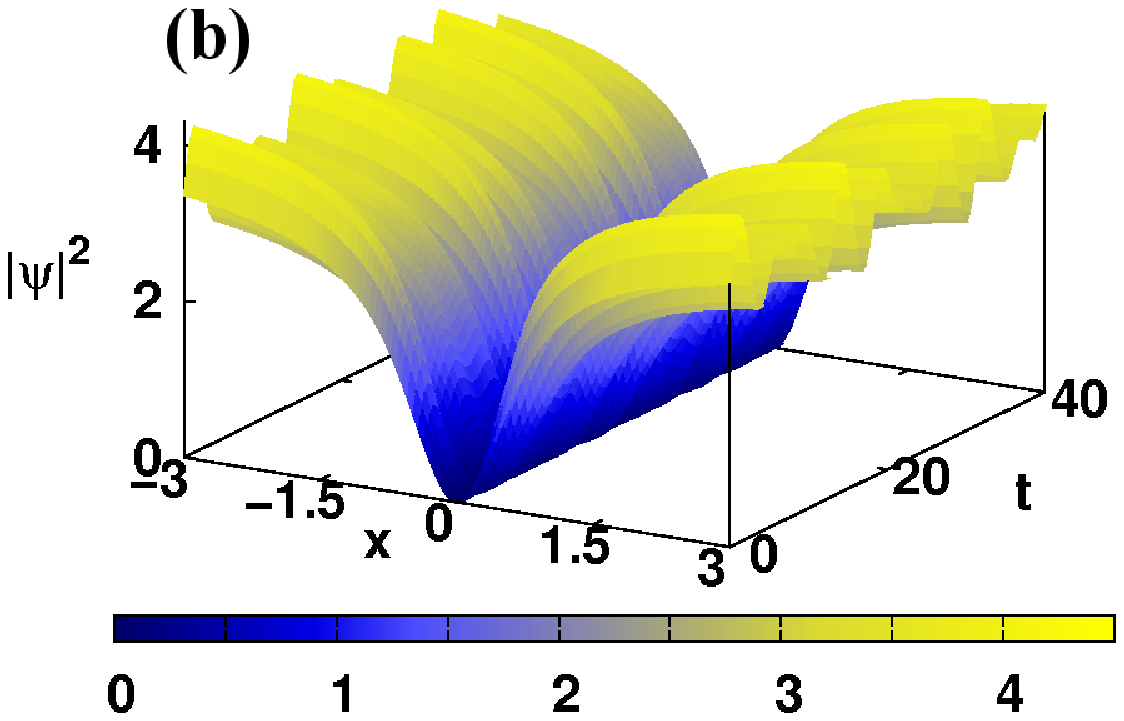}
\caption{(Color on line) Plots of $|\Psi(x,t)|^2$ with $\lambda=0.5$ for the periodic (a) or quasiperiodic (b) dark soliton, with $\alpha=0.1$ and $\beta=0$ or $\alpha=\beta=0.1$, respectively, in the case of cubic and quintic nonlinearities.}
\end{figure}
%%%%%%%%%%%%%%%%%%%%%%%%%%%%%%%%%%

We now illustrate the presence of bright and dark solitons in the case of cubic and quintic nonlinearities. In Figs.~6(a) and 6(b) we depict $|\Psi|^2$ according to \eqref{psicqb}, for $\omega(t)$ periodic and quasiperiodic, respectively. In Figs.~7(a) and 7(b) we depict $|\Psi|^2$ according to \eqref{psicqd}, for $\omega(t)$ periodic and quasiperiodic, respectively. Here we also note that the solutions evolve in time periodically or quasiperiodically, depending on the way we choose $\chi(t)$ in \eqref{chi}, with no further changes, showing that they are stable localized excitations which we name periodic and quasiperiodic bright and dark solitons for the corresponding cubic and quintic NLSE.

%%%%%%%%%%%%%%%%%%%%%%%%%%%%%%
%%%%%%%%%%%%%%%%%%%%%%%%%%%%%%
In summary, in this work we have obtained explicit soliton solutions of both the bright and dark type in the case of cubic and cubic plus quintic nonlinearities,
in the presence of potentials which oscillate in time from attractive to expulsive behavior. The procedure is valid for a diversity of nonlinearities, and the results lead to the case of periodic and quasiperiodic solutions, having bright and dark features which evolve in time maintaining their localized behavior. As we have shown explicitly, the system can support bright and dark solitons in the cases of cubic, and cubic plus quintic nonlinearities. The freedom to choose the width of the solution has allowed to describe periodic and quasiperiodic solutions, which are of current interest to the study of nonlinear excitations in condensates. Other possibilities of practical interest follow naturally.

The authors would like to thank CAPES, CNPQ and PRONEX-CNPq-FAPESQ for partial financial support.

%%%%%%%%%%%%%%%%%%%%%%%%%%%%%%%%

\end{document}